\documentstyle[emulateapj,apjfonts,psfig]{article}

\lefthead{J.~M.~Miller et al.}  \righthead{Cygnus X-1}

\received{Date}
\revised{Date}
\accepted{Date}

\journalid{vol}{date}
\articleid{1}{4}
\paperid{id}

\cpright{AAS}{1999}
\ccc{x}

\begin{document}

\title{Revealing the Focused Companion Wind in Cygnus X-1 with \textit{Chandra}}

\author{J.~M.~Miller\altaffilmark{1},
        P.~Wojdowski\altaffilmark{1},
	N.~S.~Schulz\altaffilmark{1},
	H.~L.~Marshall\altaffilmark{1},
	A.~C.~Fabian\altaffilmark{2},
	R.~A.~Remillard\altaffilmark{1},\\
	R.~Wijnands\altaffilmark{1},
	W.~H.~G.~Lewin\altaffilmark{1}
	}

\altaffiltext{1}{Center~for~Space~Research and Department~of~Physics,
        Massachusetts~Institute~of~Technology, Cambridge, MA
        02139--4307; jmm@space.mit.edu}
\altaffiltext{2}{University of Cambridge Institute of Astronomy,
        Cambridge, UK CB3 OHA}


\authoremail{jmm@space.mit.edu}

\label{firstpage}

\begin{abstract}
We have analyzed a \textit{Chandra}/HETGS spectrum of the Galactic
black hole Cygnus~X-1, obtained at a source flux which is
approximately twice that commonly observed in its persistent
low-intensity, spectrally-hard state.  We find a myriad of absorption
lines in the spectrum, including Ly-$\alpha$ lines and helium-like
resonance lines from Ne, Na, Mg, and Si.  We calculate a flux-weighted
mean red-shift of $\simeq 100$~km/s and a flux-weighted mean velocity
width of $\simeq 800$~km/s (FWHM) for lines from these elements.  We
also detect a number of transitions from Fe~XVIII--XXIV and Ni~XIX--XX
in absorption; however, the identification of these lines is less
certain and a greater range of shifts and breadth is measured.  Our
observation occurred at a binary phase of $\phi \simeq 0.76$; the
lines observed are consistent with absorption in an ionized region of
the supergiant O9.7~Iab companion wind.  The spectrum is extremely
complicated in that a range of temperatures and densities are implied.
Prior \textit{Chandra}/HETGS spectra of Cygnus~X-1 were obtained in a
similar transition state (at $\phi \simeq 0.93$) and in the low/hard
state (at $\phi \simeq 0.84$).  Considered together, these spectra
provide evidence for a companion wind that is focused as it flows onto
the black hole primary in this system.
\end{abstract}


\section{Introduction}
Cygnus~X-1 is the only known Galactic black hole which has displayed
persistent X-ray activity since the dawn of X-ray astronomy.  It is
also a rare black hole binary in that it has a high-mass companion
star (HDE~226868 --- an O9.7~Iab supergiant; Gies \& Bolton 1982,
1986).  The mass of the black hole primary is most likely $M_{1}
\simeq 10.1~M_{\odot}$, and the mass of the companion secondary is
most likely $M_{2} \simeq 17.8~M_{\odot}$ (Herrero et al. 1996).

At present, there are 14 Galactic systems with low-mass companions
($M_{2} \sim 1~M_{\odot}$) for which optical radial velocity curves
imply a primary with $M_{1} \geq 3~M_{\odot}$ (the theoretical
upper-limit for a neutron star mass; for a list of
dynamically-constrained low-mass black hole systems see
http://www.astro.uu.nl/$\sim$orosz/).  These systems are transients,
undergoing outbursts in which the X-ray luminosity may change by
factors of $10^{6}$ or more on scales ranging from days to months.  It
is logical, then, to associate the persistent nature of Cygnus~X-1
with the fact that it is a high-mass X-ray binary (HMXB).  While the
companion wind is very likely to play an important accretion role in
Cygnus~X-1 --- perhaps through a ``focused wind'' scenario (Friend \&
Castor 1982, Gies \& Bolton 1986; see below) --- numerous detections
of the disk's thermal spectrum, strong and broad Fe~K$\alpha$ emission
lines, and strong disk reflection in Cygnus~X-1 clearly demonstrate
that standard disk accretion is also important in this system.  The
relative importance of the wind and disk in driving the accretion
process and X-ray ``states'' in Cygnus~X-1 is still uncertain (for a
review of states, see Done et al. 2002; for a critical discussion see
Homan et al. 2001).

Winds from isolated massive stars may be approximated as being
spherically symmetric.  Friend \& Castor (1982) explored the geometry
of winds from massive stars in the presence of a compact object.  Due
to the compact object's gravity, continuum radiation pressure, and
centrifugal forces from binary orbital motion, it was found that winds
are likely to be strongly asymmetric if the mass donor is close to
filling its critical Roche surface.  The region of highest mass flux
is likely to be along the axis connecting the compact object and mass
donor components.  Optical spectroscopy of HDE~226868 (Cygnus X-1)
later revealed that modulations of the He II $\lambda$4686 emission
line could be described by this geometry, and that the bulk of this
emission may come from a cone ($\theta < 20^{\circ}$) along the
connecting axis (Gies \& Bolton 1986).

The sensitivity and resolution of previous X-ray observatories has
been insufficient to directly probe this focused wind geometry.
However, matters have improved greatly with the \textit{Chandra} High
Energy Transmission Grating Spectrometer (HETGS).  Marshall et
al. (2001a) discussed the detection of several highly ionized
absorption lines in a 15~ksec \textit{Chandra}/HETGS observation made
in the persistent low-luminosity, hard spectrum state (at a binary
phase of $\phi \simeq 0.84$, defining $\phi=0$ as the point at which
the companion is closest along our line of sight and based on the
ephemeris of La~Sala et al. 1998) commonly observed in Cygnus X-1.  It
is natural to associate the observed absorption features with the
companion wind as the same lines have been observed in emission in a
number of HMXBs with neutron star primaries.  Schulz et al. (2002)
report the detection of numerous emission and absorption features, and
P-Cygni-type line profiles, in a prior 15~ksec \textit{Chandra}/HETGS
observation of Cygnus X-1 (at $\phi \simeq 0.93$) made in a state
wherein the flux was approximately twice that reported by Marshall et
al. (2001a).

We observed Cygnus~X-1 with the \textit{Chandra}/HETGS for 32.1~ksec
on January 4, 2001, at a flux which was approximately twice that
commonly observed in the ``low/hard'' spectral state (we likely
observed the source in an ``intermediate'' state).  The observation
occurred at a binary phase of $\phi \simeq 0.76$ in the 5.6-day
orbital period.  The broad-band spectrum obtained in this observation
is discussed by Miller et al. (2002).  A composite Fe~K$\alpha$
emission line was revealed --- the first such composite line clearly
resolved in a binary black hole system.  The broad component may be
shaped by Doppler shifts and strong gravitational effects at the inner
edge of an accretion disk extending close to the marginally stable
circular orbit, while the neutral, narrow component may be produced in
the outer accretion disk.

Herein, we report the results of our analysis of the time-averaged
high resolution spectrum in the 6--24~\AA~ (0.5--2.0~keV) band.  While
some of the lines we detect were also found by Marshall et al. (2001a)
and Schulz et al. (2002), in this paper we present the first attempt
to systematically fit a high-resolution spectrum of Cygnus X-1 and to
measure the parameters of the strongest absorption and emission lines
in this band.  Moreover, it is clear that the spectrum we have
observed differs considerably from those reported earlier, likely
indicating that the appearance of the wind changes with orbital phase
and the intensity of the source.

\section{Observation and Data Reduction}
We observed Cygnus X-1 on 4 January 2001, from 06:03:47 to 14:59:20
(UT), for a total of 32.133~ks.  The observatory was still slewing
onto the source during the first 900~s of this observation, so we do
not include this data in our analysis.  We exclude data acquired
during a prominent 500~s dip in the X-ray lightcurve which occurs
25.7~ks into this observation.  The remaining 30.733~ks is used for
analysis.  

To prevent severe photon pile-up, the dispersed spectrum from the
HETGS was read-out with the ACIS-S array operating in
continuous-clocking (Graded-CC) mode.  To prevent telemetry saturation
and frame-dropping, the zeroth-order photons were not read-out.  This
was achieved with a 100-row blocking window.  A SIM-Z translation of +
4 mm (towards the top of the ACIS-S array) was used to limit wear and
tear on the nominal ACIS-S3 aimpoint.  A Y-coordinate translation of
-80 arcseconds (towards S4) was used to place as much of the iron line
region on S3 as possible.

At the time we started our analysis, the standard CIAO processing
tools were unable to handle data taken in this mode.  Therefore, we
developed a robust set of custom processing routines.  This processing
method is described in great detail in a paper reporting results from
a Chandra/HETGS observation of the Rapid Burster in outburst (Marshall
et al. 2001b).  All aspects of how the grating orders and background
were selected, and how the data is corrected for the instrument
response, are exactly those previously reported.  The only important
difference between that analysis, and our analysis of Cygnus X-1 is
the blocking of the zeroth-order photons.  By examining the location
of the neutral Si absorption edges (due to the Si-based CCDs) in
opposite grating orders in raw counts space, we were able to derive a
zeroth-order position.  This position was fine-tuned by fitting the
most prominent absorption lines in opposite grating orders and
requiring that the centroid wavelengths agree to within 0.05\%
uncertainty, which matches the HETG calibration uncertainty (Canizares
et al., 2002, in prep.).  We are confident that our wavelength
calibration is equivalent to that for standard observing modes.

Fits to the high-resolution data were made using ISIS version 0.9.37
(Houck \& Denicola 2000).  Systematic errors were not added for our
analysis of the narrow spectral features (in this mode, flux
uncertainties across the full HETGS band may be 5\%; Marshall et
al. 2001b).  We estimate that the effects of photon pile-up in our
observation to be negligible.  We also note instrumental features near
6.71\AA~ (a single-bin ``line'' at HEG resolution near the Si K edge)
and a broader feature at 7.96\AA~ due to the instrumental Al K edge
(Marshall 2002, in prep.).

\section{Analysis and Results}
The most striking result of this observation is the wealth of
absorption lines found in the 6--24\AA~ range (concentrated in the
6--15\AA~band, see Figure 1 and Figure 2).  Our fits to the absorption
and emission features are listed in Table 1.  Each line is significant
at or above the 3$\sigma$ level of confidence unless noted explicitly
as an upper-limit; 34 lines are detected and resolved.

In analyzing the high resolution spectrum in the 6--15\AA~range, we
combine the HEG orders and MEG orders.  These spectra were fit jointly
with local single-component models for the continuum in contiguous
3\AA~ segments.  The HEG range does not extend to wavelengths longer
than 14\AA, so only the MEG spectrum is considered between 14--24\AA.
All of the lines in the high resolution spectra are fit with Gaussian
profiles.  We fit only the strongest features, with the requirement
that the feature be evident in each of the four first-order spectra
between 6--14\AA, and in both first-order MEG spectra above 14\AA.  In
cases where nearby line features might cause blending, the width is
fixed to $\sigma=0.01$ \AA~ -- the maximum line resolution of the HEG.
For such lines, the FWHM is 0.0236~\AA~ -- just broader than the
maximum line resolution of the MEG.  This width is generally
consistent with the lines in Table 1 for which the width was allowed
to vary.

Identifications for the line features were made by assuming the
transition with highest oscillator strength in the wavelength range of
a given feature, using the line lists of Verner, Verner, \& Ferland
(1996) and Mewe, Gronenschild, \& van den Oord (1985).  It should be
noted that many lines identified in this work may be blends, as Fe
L~shell transitions are present throughout this range.  In examining
Figure 1 and Figure 2, it is clear that a myriad of weaker lines exist
in this spectrum which we have not identified here; the majority of
these features are likely absorption lines from the Fe L-shell.

The column densities of observed lines are calculated using the
relation:\\
\begin{center}
(1)~~ $W_{\lambda} = \frac{\pi e^{2}}{m_{e} c^2} N_{j} \lambda^{2} f_{ij} =
8.85 \times 10^{-13} N_{j} \lambda^{2} f_{ij}$\\
\end{center}
\noindent where $N_{j}$ is the column density of a given species,
$f_{ij}$ is the oscillator strength, $W_{\lambda}$ is the equivalent
width of the line, and $\lambda$ is the wavelength of the line in cm
units (Spitzer 1978).  We take $f_{ij}$ values for all resonant lines
from Verner et al. (1996), and those for other lines from Mewe et
al. (1985).  In using this equation, we have implicitly assumed that
we are on the linear part of the curve of growth and that the lines
are not significantly saturated.

\subsection{Absorption in the Interstellar Medium}
We clearly detect the neutral K-shell absorption edges of O, Ne, Mg,
and Si, and Fe-L3 and Fe-L2 edges in the 6--24\AA~ range (see Figure
1, Figure 2).  Following the procedure of Schulz et al. (2002) in a
previous analysis of the ISM absorption edges in a \textit{Chandra}
spectrum of Cygnus~X-1, we fit the K-shell edges with a simple
step-function edge model based on the photoelectric cross sections of
Henke, Gullikson, \& Davis (1993) and elemental abundances relative to
solar as per Morrison \& McCammon (1983).  We fit the Fe~L edges with a
model based on the cross sections of Kortright \& Kim (2000).

We find that the wavelength, depth ($\tau$), and abundances (relative
to solar) implied by the K-shell edges are fully consistent with the
values reported by Schulz et al. (2002) based on measurements at $\phi
\simeq 0.93$, and we refer the reader to that work for a full
discussion.  The depths of the Fe~L edges are also consistent; again
the edges are found at 17.51\AA~ and 17.22\AA~ (rather than 17.56\AA~
and 17.23\AA, perhaps indicating non-metallic contributions).

The $1s-2p$ transition from neutral (atomic) O is detected at $23.50
\pm 0.02$~\AA~ with a flux of $-2.0 \pm 0.5 \times 10^{-4}~ {\rm
ph}~{\rm cm}^{-2}~{\rm s}^{-1}$.  The best-fit width of this line is
quite large: ${\rm FWHM} = 0.18 \pm 0.04$~\AA.  This line is likely
saturated, however, so the flux and FWHM measurements are likely
under-estimated and over-estimated, respectively.

\subsection{Hydrogenic and Helium-like Ne, Na, Mg, and Si Lines}
We clearly detect the Ly-$\alpha$ lines of Ne, Na, Mg, and Si in
absorption (see Figure 1 and Figure 2, and Table 1).  The Ne series is
seen in absorption from Ly-$\alpha$ to Ly-$\delta$.  The helium-like
resonance ($r$) lines of these elements are also clearly detected in
absorption.  We also fit the inter-combination ($i$) and forbidden
($f$) lines from these elements, which are observed to be in emission.
Due to a nearby error in the the instrument response we tie the Si $f$
emission line width and strength to those values measured for the $r$
line.  The Si $i$ line is not clearly detected, and measurements
should be regarded as upper-limits.  Similarly, the Ne $f$ line
parameters should be regarded as upper-limits.

The measured equivalent widths of the Ne Ly series do not decrease
steadily in strength from Ly-$\alpha$ to Ly-$\delta$ as one would
expect from the oscillator strengths for these lines (see Table 1).
This suggests that the Ne Ly series absorption lines may be partially
saturated, or that regions of different density are seen
simultaneously (also supported by the inconsistent FWHM values we
measure).  It is also possible that unresolved Fe L-shell transitions
contribute to the absorption, and that the lines are actually blends.
Inconsistent line strengths and column densities are also measured
from the helium-like Ne, Na, Mg, and Si $r-i-f$ triplets, though some
of these are poorly constrained and/or upper-limits (see Table 1).

Including the helium-like and hydrogenic lines from Ne, Na, Mg, and Si
for which the FWHM could be constrained, we calculate a flux-weighted
mean FWHM of $\simeq 800$~km/s.  For the same lines, we calculate a
flux-weighted mean red-shift of $\simeq 100$~km/s.

\subsection{Highly-Ionized Fe and Ni Lines}

A number of lines which we identify with transitions from
Fe~XVIII--XXIV are also detected, as well as lines which may be from
Ni~XIX and Ni~XX.  It must be noted that the identification of these
line is less certain than those from lower-Z elements, and this may
account for a greater spread in the implied bulk velocity shifts and
velocity widths measured for these lines.

We used a boot-strap method to identify lines in this spectrum.  The
Ly series lines from lower-Z elements are fewer than the numerous
transitions from the Fe~L-shell, and spaced at regular intervals.
Most of those lines are measured to have slight red-shifts, or to be
consistent with zero bulk velocity.  In examining other lines, then,
we made the assumption that observed lines should lie relatively close
to their rest-frame wavelengths.  However, some are observed at
blue-shifts (e.g., Fe~XVIII at 14.257\AA~ -- blue-shifted by
1000~km/s).  The identifications given to these lines are tentative,
but reflect the transition with highest oscillator strength in the
given region.

As with the lower-Z lines, the measured strengths and implied column
densities for multiple lines from the same species are not consistent.
Again, in such cases it is possible that the given line is actually a
blend.

\subsection{Other Line Features}

We note a pair of marginal emission features at 21.8~\AA~ and
21.9~\AA.  Although the sensitivity of the spectrum is limited in this
range, modeling these lines improves the fit in this region.  The line
at 21.8~\AA~ can be identified as the the O~VII $i$ line.  In general,
this triplet line is only expected in isolation if produced in a very
dense region.  For this reason, it is tempting to associate this line
with the accretion disk.  It is possible that the line at 21.9~\AA~ is
a second O~VII $i$ line, part of a Doppler-shifted pair that are at an
overall red-shift.  In this case, the overall red-shift is $\simeq
700$~km/s and the separation implies a Doppler shift of $\simeq
1400$~km/s.  While the Doppler shift is plausibly associated with the
outer accretion disk, the overall red-shift of the pair is not.

\subsection{Constraints on the Companion Wind at $\phi \simeq 0.76$}

While the observed spectrum is clearly very complicated, some basic
inferences can be made.  Miller et al. (2002) find that the
broad-band spectrum from this observation of Cygnus~X-1 is strongly
dominated by a $\Gamma \sim 1.8$ power-law component, and that the
source was at a luminosity of $L_{X} \sim 1.0 \times 10^{37}~{\rm
erg}~{\rm s}^{-1}$ (0.65--10~keV).  Assuming a $\Gamma=2.0$ power-law
input flux and overall luminosity of $L_{X}=10^{37}$ ergs/s in the
soft X-ray band, Kallman \& McCray (1982) have calculated the
temperatures at which absorption lines from given elemental species
will be at peak strength.  Here, we list these temperatures for select
species: for Ne~IX, log(T)$=$4.5-5.0; for Ne~X, log(T)$=$5.0-5.5; for
Si~XII, log(T)$=$5.2-5.7; and for Si~XIV, Fe~XVIII, and Fe~XIX,
log(T)$=$5.5-6.0.  Thus, the absorption features we detect arise in
regions that span over an order of magnitude in temperature.  Given
the range in ionization states observed and temperatures implied, it
is likely that the gas in the line of sight cannot be described by a
single ionization parameter ($\xi = L_{X} / n r^{2} = L_{X} / {\rm
N}_{H} r$, where $L_{X}$ is the luminosity, $n$ is the hydrogen number
density, ${\rm N}_{H}$ is the hydrogen equivalent column density, and
$r$ is radius).  If $L_{X}$ may be assumed to be relatively steady,
then $n r^{2}$ may vary significantly along our line of sight.

A number of additional constraints can be derived by considering the
ionization fractions expected in this regime.  Kallman \& Bautista
(2001) have calculated the ionization fractions expected for a gas
with typical wind characteristics, again assuming a $\Gamma=2.0$
power-law input flux and overall luminosity of $L_{X}=10^{37}$ ergs/s
in the soft X-ray band.  The hydrogen equivalent column density for
resonance absorption lines is given by ${\rm N}_{H,Z} = {\rm N}_{Z} /
(F_{Z} \times A_{Z})$ (where ${\rm N}_{H,Z}$ is the hydrogen
equivalent column density for a given resonance line, ${\rm N}_{Z}$ is
the column density for that line as noted in Table 1, $F_{Z}$ is the
ionization fraction for the given ion species, and $A_{Z}$ is the
solar abundance of the given element relative to H).

We find that the hydrogen equivalent column densities of helium-like
and hydrogenic resonance lines from given single elements are broadly
self-consistent for those elements: ${\rm N}_{H, Ne~IX, X} \sim 7.2
\times 10^{20}~ {\rm cm}^{-2},~ {\rm N}_{H, Mg XI, XII} \sim 14.5
\times 10^{20}~ {\rm cm}^{-2},~ and {\rm N}_{H, Si~XIII, XIV} \sim 7.4
\times 10^{22}~ {\rm cm}^{-2}$.  As noted previously, the Ne~Ly series
is likely partially saturated.  Performing the above exercise using
the Ne~Ly-$\gamma$ and Ne~Ly-$\delta$ lines gives ${\rm N}_{H, Ne~X}
\sim 4.3 \times 10^{21}~ {\rm cm}^{-2}$; this likely serves as a
lower-limit estimate on the hydrogen equivalent column density of
Ne~X.  With the exception of the Fe~XIX lines at 14.93\AA~ and
14.97\AA, the hydrogen equivalent column densities for the ionized Fe
species (Fe~XVIII--XXIV) are below $10^{23}~ {\rm cm}^{-2}$.

Comparing again to Kallman \& Bautista (2001), the absorption region
can be described by ionization parameters between ${\rm log}(\xi) =
2-3$, corresponding to distances from the central ionizing source of
$(5-9) \times 10^{10}$ cm.  If we assume that the binary separation of
Cygnus~X-1 is $40~R_{\odot}$, and that the radius of HDE~226868 is
$20~R_{\odot}$ (consistent with the parameters derived by LaSala et
al.\ 1998), then the black hole is approximately $1.4 \times
10^{12}$~cm from the companion surface.  The distances implied by the
range of ionization parameters are far below this separation.  This
disparity can be seen a second way: if the companion wind is
spherically symmetric and our line of sight extends through a distance
similar to the black hole companion-surface separation, assuming
typical stellar wind densities (${\rm log}[n] = 10-11$) we would
expect to measure hydrogen equivalent column densities of $1.4 \times
10^{22-23}~ {\rm cm}^{-2}$.  In contrast, the strong majority of the
estimates we have obtained are below this range.

The $f$/$i$ line intensity ratio for a given helium-like element can
be used to measure the density of emitting material.  Porquet \& Dubau
(2000) have calculated densities as a function of the gas temperature
for a number of elements, including Ne, Mg, and Si among the lines we
have detected.  Dividing the column density measurements obtained via
equation (1) by these densities, then, an estimate of the physical
depth of the absorbing gas along our line of sight may be obtained.
We assume the temperatures of peak absorption for Ne IX, and Si XIII
from Kallman \& McCray (1982), and assume the appropriate temperature
for Mg XI lies between these values.  As the $i$ and $f$ lines we
observe are weak, their intensities may be regarded as upper-limits.
The linear depth estimates we obtain via the calculations of Porquet
\& Dubau are not very constraining: we find lower-limits of only
10-100~m.

Finally, we do not find convincing evidence of P-Cygni line features
like those reported by Schulz et al. (2002) at $\phi \simeq 0.93$.
This fact may indicate that the appearance of such line profiles is
phase-dependent; it is also possible that the appearance of nearby
emission and absorption features is coincidental, as suggested by the
authors.

\section{Discussion}
We have performed the first systematic fits to the line-rich
low-energy spectrum of Cygnus X-1.  The temperatures, velocities, and
densities implied by the weak, narrow lines in the spectrum are
consistent with absorption in an ionized portion of the companion
wind.  The spectrum we have observed from Cygnus X-1 is dominated by
absorption lines, however, which stands in strong contrast to the
spectra observed from other HMXBs.  \textit{ASCA} and \textit{Chandra}
observations of Vela X-1 (Sako et al. 1999, Schulz et al. 2002b) and
Cen X-3 and (Wojdowski, Liedahl, \& Sako 2001) reveal emission line
spectra due to recombination in a photoionized companion wind (the
emission lines are most prominent in eclipse, but persist outside of
eclipse).  A recent \textit{Chandra} observation of Cygnus X-3
revealed many of the same strong emission lines independent of orbital
phase (Paerels et al. 2000).

While the spectrum of Cen X-3 can be modeled in terms of a
spherically-symmetric wind centered on the neutron star (Wojdowski,
Liedahl, \& Sako 2001), the absorption spectrum of Cygnus X-1 may
require dense material preferentially along the line of sight.  A
relative lack of material outside of the line of sight is also
required to explain the absence of resonance emission lines excited by
photoionization from the central accretion engine.  The inclination of
Cygnus X-1 is rather low ($\theta \simeq 35^{\circ}$; Gies \& Bolton
1986), so it is unlikely that the absorption region can be associated
with the outer accretion disk or a warm (few keV) disk atmosphere.  As
optical observations of Cygnus X-1 (Gies \& Bolton 1986) have already
provided evidence for a focused wind geometry, we suggest that such a
geometry might also explain the X-ray spectrum we have observed.  In
the discussion below, we first examine the evidence for a focused wind
from this and other recent \textit{Chandra} observations of Cygnus
X-1.  We then address the implications of the neutral absorption
features and the impact of this observation on our understanding of
Cygnus X-1.

\subsection{Evidence for a Focused Wind Geometry}
Kallman \& Bautista (2001) have calculated ionization fractions and
ionization parameters for a gas with properties consistent with
stellar winds, for an incident spectrum similar to that measured here
(Miller et al. 2002).  The ionization parameters consistent with the
observed absorption spectrum (${\rm log}(\xi) = 2-3$) imply distances
between the source and absorbing gas ($5-9 \times 10^{10}$~cm) that
are far less than the distance between the black hole and the
companion surface ($\sim 1.4 \times 10^{12}$~cm; LaSala et al.\ 1998).
Moreover, the estimates we obtain for the equivalent neutral hydrogen
column densities for various elements generally lie well below
expectations for a spherically symmetric wind (assuming that our line
of sight through the wind is comparable to the separation between the
black hole and companion surface).

An examination of the velocities implied by the absorption lines also
supports a focused wind geometry.  For the Ly-$\alpha$ and helium-like
resonance lines from Ne, Mg, Na, and Si, we measure a flux-weighted
mean red-shift of $\simeq$100~km/s and flux-weighted mean FWHM of
$\simeq$800~km/s (again, our observation occurred at $\phi \simeq
0.76$).  Marshall et al. (2001a) report a number of absorption lines
with a mean red-shift of $\sim$450~km/s and a typical FWHM of
$\sim$300~km/s at $\phi \simeq 0.84$.  This is consistent with the
expectation that the a focused flow should be largely transverse to
the line of sight at intermediate phase points, and have a greater
component along the line of sight at other phases.

Schulz et al. (2002) report marginal evidence for ionized Fe
transitions with P-Cygni-type line profiles at $\phi = 0.93$.  This
may indicate that the wind from the companion surface opposite to the
X-ray source may have a significant radial component.  The absence of
such line profiles at more intermediate phases (this work; Marshall et
al. 2001a) suggests that the wind does not have a radial component of
typical strength at phases away from conjunction.

We suggest that when considered together, these spectra may provide
the first direct evidence in X-rays for a focused wind geometry in
Cygnus X-1 (see Figure 3 for a possible geometry).  The column
densities measured from absorption lines are smaller than expected for
a spherically symmetric wind geometry.  At an intermediate phase point
the flow is mostly transverse to the line of sight; at points closer
to superior conjunction, the flow has smaller transverse velocities
and higher velocities parallel along the line of sight.  Finally, very
close to superior conjunction, the wind from the face of the companion
opposite to the X-ray source may have a radial component; indeed
P-Cygni profiles may have been observed by Schulz et al. (2002a).

A caveat is that the observation reported on by Marshall et
al. (2001a) occurred in the low/hard state typically observed in
Cygnus X-1, while this observation and that discussed by Schulz et
al. (2002a) were made during transitional states with higher fluxes.
At present, it is not clear how the source state and wind are related.
Moreover, all three observations were separated by several months, and
the nature of the wind may vary with the long-term period in this
system (e.g., Brocksopp et al. 1999).  The observation of narrow,
blue-shifted absorption lines from low-Z elements near to inferior
conjunction and broadened lines near $\phi = 0.25$ would strengthen
the evidence for a focused wind accretion geometry.  Future
observations with the \textit{Chandra}/HETGS (or
\textit{XMM-Newton}/RGS) achieving sensitivities similar to those
observations considered here will be required for this purpose.

\subsection{Neutral Absorption Edges, and the Atomic O $1s-2p$
Absorption Line}

We measure absorption edge depths from neutral atoms in the
interstellar medium that are consistent with those reported by Schulz
et al. (2002).  This suggests that any cold absorbing gas intrinsic to
Cygnus X-1 does not vary with orbital phase.  Although the atomic O
$1s-2p$ line is saturated (and its profile therefore distorted) in
both observations, the width we measure (FWHM$ = 0.18 \pm 0.04$\AA) is
roughly twice the width measured previously.  

The width of the O $1s-2p$ lines measured in observations of Cygnus
X-1 are factors of 10--20 higher than the width of the same line
measured in a \textit{Chandra}/LETGS spectrum of the neutron star
4U~0614$+$091 (Paerels et al. 2001).  Although the lines in Cygnus X-1
are saturated, this disparity is likely larger than can be attributed
solely to distortion of the line Cygnus X-1 profile.  It is possible
that the variation can be accounted for by different turbulent
velocities in the neutral oxygen intrinsic to each system.  However,
due to the ionizing X-ray flux from accretion onto the compact object
in these systems, little neutral oxygen is expected locally.  We
suggest that the atomic O $1s-2p$ line, then, may serve as a probe of
turbulent velocities in the ISM along different lines of sight.
Studies of the Galactic H~I power spectrum which suggest that column
density variations due to cold gas may be dominated by localized
velocity field fluctuations (Dickey et al. 2001).  At present, this is
a relatively unexplored regime in X-ray astronomy but well within the
reach of current observatories.

\subsection{Understanding the Accretion Geometry in Cygnus X-1}
Future analysis and observations will achieve a better understanding
of the degree to which the wind in Cygnus X-1 may be focused, and its
temperature and density characteristics.  In the near future, then, it
may be possible to constrain the mass accretion rate onto the black
hole via focused wind accretion.  Comparisons to the observed X-ray
luminosity will then indirectly provide a constraint on the mass
delivered to the black hole via disk accretion.

Constraints of this kind are important for a number of reasons.
Cygnus X-1 often serves as a standard for testing models for accretion
flow geometries and state transitions based on the mass accretion rate
(such as advection-dominated accretion flow models, or ``ADAFs''; see,
e.g., Esin et al. 1998).  This source is also a testbed for X-ray
``reflection'' models (see, e.g., Ross, Fabian, \& Young 1999), which
suggest that the innermost accretion flow geometry in Cygnus X-1 may
be very similar to that in some AGN.  In both cases, a central source
of hard X-rays (a ``corona'') is important, as is an accretion disk
(some reflection models suggest that an ionized transition layer may
lie on top of the accretion disk).  Neither family of models has
considered the role of the companion wind, or the nature of disk-wind
or corona-wind interactions.

Although the timing properties of Cygnus X-1 are well-studied (see,
e.g., Pottschmidt et al. 2002), quasi-periodic oscillations (QPOs)
have never been observed at high frequencies (``high'' is somewhat
arbitrary, $\nu > 30~{\rm Hz}$ is a reasonable distinction).  Such
QPOs have been very useful in constraining black hole spin in systems
such as GRO~J1655$-$40 (Strohmayer 2001); in contrast spin can only be
investigated in Cygnus X-1 via spectroscopy (e.g., Miller et
al. 2002).  It is possible that the companion wind in Cygnus X-1 acts
to dampen high-frequency signals preferentially; when a better
understanding of the wind geometry and density is achieved this
possibility can be addressed.

At present, Cygnus X-1 is the only Galactic HMXB for which a black
hole primary is required.  LS~5039 may provide an interesting
comparison: the companion is an O6.5 V(f) supergiant (which does not
fill its critical Roche radius), the orbital period is 4.1 days, and
its distance is likely $\sim$3.1~kpc (McSwain \& Gies 2002).  Our
preliminary analysis of archival \textit{ASCA} data suggests that
LS~5039 is very weak in X-rays relative to Cygnus X-1: $L_{X} < 2
\times 10^{34}~ {\rm erg}~ {\rm s}^{-1}$ (0.5--10.0 keV).  This may
underscore the importance of disk accretion via Roche-lobe overflow in
Cygnus X-1.  At present, the nature of the compact object in LS~5039
is not known.  Even if the primary in LS~5039 is a neutron star, this
system bears important similarities to Cygnus X-1, and we look forward
to spectroscopy of this source with \textit{Chandra} and
\textit{XMM-Newton}.

\section{Acknowledgments}
We wish to thank \textit{Chandra} Director Harvey Tananbaum, and the
\textit{Chandra} staff for executing this observation and their help
in processing the data.  We wish to recognize John Houck and David
Huenemoerder for their insights, and the CXC/ISIS team.  We thank
Claude Canizares, Michael Nowak, and Bryan Gaensler for helpful
insights.  We acknowledge Adrienne Juett for sharing the specialized
Fe~L absorption model.  RW was supported by NASA through Chandra
fellowship grants PF9-10010, which is operated by the Smithsonian
Astrophysical Observatory for NASA under contract NAS8--39073.  WHGL
gratefully acknowledges support from NASA.  This research has made use
of the data and resources obtained through the HEASARC on-line
service, provided by NASA-GSFC.

\clearpage


\begin{table}[t]
\caption{Prominent Emission \& Absorption Features}
\begin{scriptsize}
\begin{center}
\begin{tabular}{lllllllllll}
\multicolumn{3}{l}{Ion and} & Theor. & Meas. & Shift &
\multicolumn{2}{c}{FWHM} & W & Flux & N$_{\rm Z}$ \\
\multicolumn{3}{l}{Transition} & (\AA) & (\AA) & (km/s) & ($10^{-2}$\AA) &
(km/s) & (m\AA) & ($10^{-4}$~ph/cm$^{2}$/s) & ($10^{15}~{\rm cm}^{-2}$) \\
\tableline

\multicolumn{3}{l}{Si XIV 1s-2p} & 6.1822 & 6.1808(9) & $+$20(50) &
$2.6^{+0.4}_{-0.2}$ & $1300^{+200}_{-100}$ & $-5.0(4)$ & $-7.7(6)$ & 36(3)\\

\multicolumn{3}{l}{Si XIII r} & 6.6408 & $6.652^{+0.001}_{-0.003}$ &
$-490^{-80}_{+120}$ & $1.8^{+0.8}_{-0.4}$ & $800^{+400}_{-200}$ &
$-2.0(4)$ & $-3.0^{-0.7}_{+0.6}$ & $6.6^{+1.6}_{-1.3}$\\

\multicolumn{3}{l}{Si XIII i~$\dag$} & 6.690 & 6.690 & -- & 2.36 & 1100
& $0.5^{+0.3}_{-0.4}$ & $0.7^{+0.6}_{-0.5}$ & 1.6(1.3) \\

\multicolumn{3}{l}{Si XIII f~$\dag\dag$} & 6.740 & 6.7366(3) & 150(20)
& 1.8 & 800 & 2.0 & 3.0 & $6.6^{+1.6}_{-1.3}$\\

\multicolumn{3}{l}{Mg XI} & 7.8505 & 7.853(3) &
$-100^{-100}_{+90}$ & $1.5^{+0.4}_{-0.6}$ & 600(200) & $-0.9(3)$ &
$-1.4(4)$ & 11(3) \\

\multicolumn{3}{l}{(1s$^{2}$-1s3p)} & ~ & ~ & ~ & ~ & ~ & ~ & ~ & ~ \\

\multicolumn{3}{l}{Mg XII 1s-2p} & 8.4210 & 8.423(1) & $-70(40)$ &
2.6(3) & 900(100) & $-4.3(3)$ & $-6.2(5)$ & 16(1) \\

\multicolumn{3}{l}{Ni XIX} & 8.7250 & 8.720(1) &
160(30) & $0.09^{+0.03}_{-0.09}$ & $300^{+400}_{-300}$ &
$-1.1^{-0.3}_{+0.2}$ & -1.6(3) & 24(5) \\

\multicolumn{3}{l}{(2p$^{6}$-2p$^{5}$6d)} & ~ & ~ & ~ & ~ & ~ & ~ & ~ & ~ \\

\multicolumn{3}{l}{Mg XI r} & 9.1688 & 9.174(2) & $-150^{-80}_{+60}$ &
$0.8^{+0.3}_{-0.2}$ & $650^{+120}_{-140}$ & $-2.2^{-0.5}_{+0.4}$ &
$-3.0^{-0.7}_{+0.6}$ & $3.9^{+1.0}_{-0.7}$ \\

\multicolumn{3}{l}{Mg XI i} & 9.230 & 9.229(2) & $50^{+50}_{-70}$ &
$1.7^{+0.3}_{-0.5}$ & 600(200) & 2.5(4) & 3.5(6) & 4.6(8) \\

\multicolumn{3}{l}{Mg XI f} & 9.319 & $9.317^{+0.002}_{-0.004}$ &
$-200^{-70}_{+130}$ & 2.36 & 800 & 2.8(4) & $3.8^{+0.5}_{-0.6}$ & 4.9(7)\\

\multicolumn{3}{l}{Ne X 1s-5p} & 9.4807 & 9.476(2) & $140^{+70}_{-50}$
& $1.1^{+0.7}_{-0.4}$ & 400(200) & $-1.5(4)$ & $-1.9^{-0.6}_{+0.5}$ & 140(40) \\

\multicolumn{3}{l}{Ne X 1s-4p} & 9.7082 & $9.716^{+0.001}_{-0.002}$ &
$-240^{-40}_{+50}$ & 2.6(3) & $800^{+100}_{-200}$ &
$-4.2^{-0.4}_{+0.5}$ & $-5.8^{-0.5}_{+0.7}$ & 170(20)\\

\multicolumn{3}{l}{Fe XX} &
9.991 & $10.004^{+0.001}_{-0.002}$ & $-400^{-50}_{+40}$ & 2.36 & 700 & $-3.7^{-0.3}_{+0.4}$
& $-4.9^{-0.4}_{+0.6}$ & 33(3) \\

\multicolumn{3}{l}{(2s$^{2}$2p$^{3}$-2p$^{2}(^{3}{\rm P})$4d)} & ~ & ~ & ~ & ~ & ~ & ~ & ~ & ~ \\

\multicolumn{3}{l}{Na XI 1s-2p} & {10.0250} &
$10.046^{+0.002}_{-0.003}$ & $-600^{-60}_{+90}$ & 2.36 & 700 &
$-3.3^{-0.3}_{+0.4}$ & $-4.3^{-0.4}_{+0.5}$ & 9(1) \\

\multicolumn{3}{l}{Fe XX} &
10.120 & 10.132(1) & $-360(30)$ & 0.8(4) & $-$230(130) & $-1.2(3)$ & $-1.6(3)$ & $7.0^{+1.8}
_{-1.6}$ \\ 

\multicolumn{3}{l}{(2s$^{2}$2p$^{3}$-2p$^{2}(^{3}{\rm P})$4d)} & ~ & ~ & ~ & ~ & ~ & ~ & ~ & ~  \\

\multicolumn{3}{l}{Ne X 1s-3p} & 10.2389 & 10.244(2) & $-150(60)$ &
1.2(3) & 350(90) & $-1.5^{-0.2}_{+0.4}$ & $-1.9^{-0.3}_{+0.5}$ & $20^{+4}_{-5}$ \\

\multicolumn{3}{l}{Fe XXIV} & 10.619 & 10.628(1)
& $-250(30)$ & 2.36 & 670 & $-5.5(4)$ & $-6.7^{-0.6}_{+0.7}$ & 21(2) \\

\multicolumn{3}{l}{(1s$^{2}$2s-1s$^{2}$3p)} & ~ & ~ & ~ & ~ & ~ & ~ & ~ & ~ \\

\multicolumn{3}{l}{Fe XXIII} & 10.981 &
$10.985^{+0.003}_{-0.002}$ & $-110^{-80}_{+60}$ & 2.36 & 650 &
$-3.5^{-0.3}_{+0.6}$ & $-4.1^{-0.3}_{+0.7}$ & $4.7^{+0.4}_{-0.8}$\\

\multicolumn{3}{l}{(2s$^{2}$-2s3p)} & ~ & ~ & ~ & ~ & ~ & ~ & ~ & ~ \\

\multicolumn{3}{l}{Na X r} & 11.0027 & $11.015^{+0.002}_{-0.003}$ &
$400^{+80}_{-60}$ & 2.36 & 640 & $-3.3^{-0.4}_{+0.5}$ &
$-3.8^{-0.6}_{+0.5}$ & 4.1(6) \\

\multicolumn{3}{l}{Na X i} & 11.080 & $11.086^{+0.003}_{-0.006}$ &
$-160^{-80}_{+160}$ & 2.36 & 640 & $2.6(5)$ & 3.0(6) & 3.2(6) \\

\multicolumn{3}{l}{Na X f} & 11.190 & 11.190(5) & 0(130) & 2.36 & 630
& $1.7^{+0.4}_{-0.6}$ & $2.0^{+0.5}_{-0.7}$ & 2.1(5) \\

\multicolumn{3}{l}{Fe XXII} &
11.440 & $11.429^{+0.002}_{-0.001}$ & $290^{+30}_{-50}$ & 2.36 & 620 &
$-4.6^{-0.4}_{+0.6}$ & $-5.1^{-0.5}_{+0.6}$ & $12.0^{+1.2}_{-1.4}$\\

\multicolumn{3}{l}{(2s$^{2}$2p-2s2p($^{3}{\rm P}^{0}$)3p)} & ~ & ~ & ~ & ~ & ~ & ~ & ~ & ~ \\

\multicolumn{3}{l}{Fe XXII} & 11.770 & 11.773(1) & $-80(30)$ & 2.36 &
600 & $-9.2^{-0.6}_{+0.8}$ & $-9.9^{-0.7}_{+0.8}$ & $11.0^{+0.8}_{-1.0
}$ \\

\multicolumn{3}{l}{(2s$^{2}$2p-2s$^{2}$3d)} & ~ & ~ & ~ & ~ & ~ & ~ & ~ & ~ \\

\multicolumn{3}{l}{Ni XX} & 11.980 &
11.981(2) & $-30(50)$ & 2.36 & 600 & $-6.5^{-0.6}_{+0.5}$ &
$-6.9^{-0.6}_{+0.5}$ & $12.8^{+0.5}_{-0.9}$\\

\multicolumn{3}{l}{(2s$^{2}$2p$^{5}$-2s2p$^{5}$(3P)3p)} & ~ & ~ & ~ & ~ & ~ & ~ & ~ & ~ \\

\multicolumn{3}{l}{Ne X 1s-2p} & 12.1339 & $12.138^{+0.001}_{-0.002}$
& $-100^{-30}_{+50}$ & $2.6^{+0.4}_{-0.3}$ & $640^{+100}_{-70}$ &
$-10.0^{-1.5}_{-1.1}$ & $-9.0^{-1.0}_{+1.0}$ & 19(2) \\

\multicolumn{3}{l}{Fe XXI} & 12.285 &
12.292(1) & $-170(20)$ & 2.36 & 580 & $-13.1^{-0.7}_{+1.1}$ &
$-11.2^{-0.6}_{+0.9}$ & $10.1^{+0.5}_{-0.8}$ \\

\multicolumn{3}{l}{(2s$^{2}$2p$^{2}$-2s$^{2}$2p3d)} & ~ & ~ & ~ & ~ & ~ & ~ & ~ & ~  \\

\multicolumn{3}{l}{Fe XX} &
12.820 & 12.839(2) & $-450(50)$ & $3.7^{+0.4}_{-0.3}$ & $870^{+90}_{-70}$ &
$-18^{-1.7}_{+1.6}$ & $-13.5(1.2)$ & 13(1) \\

\multicolumn{3}{l}{2s$^{2}$2p$^{3}$-2p$^{2}$($^{3}$P)3d} & ~ & ~ & ~ & ~ & ~ & ~ & ~ & ~ \\

\multicolumn{3}{l}{Ne IX r} & 13.4471 & $13.448^{+0.007}_{-0.003}$ &
$0^{-160}_{+70}$ & 2.36 & 530 & $-7.2^{-1.3}_{+1.4}$ &
$-4.6^{-0.9}_{+1.0}$ & 6.2(1.2) \\

\multicolumn{3}{l}{Ne IX i} & 13.550 & 13.561(4) & $-240(90)$ & 2.36 &
520 & $11.6^{+1.6}_{-1.5}$ & $7.2^{+1.0}_{-0.9}$ & 10(1) \\

\multicolumn{3}{l}{Ne IX f $\dag$} & 13.700 & 13.700 & -- & 2.36 & 520 &
$4.5^{+0.6}_{-3.2}$ & $2.7^{+0.4}_{-1.9}$ & $3.8^{+0.5}_{-2.7}$ \\

\multicolumn{3}{l}{Fe XVIII} &
14.257 & $14.209^{0.001}_{-0.003}$ & $1000^{+60}_{-20}$ &
$0.7^{+1.0}_{-0.6}$ & $150^{+200}_{-140}$ & $-7.0^{-2.4}_{-1.7}$ & $-3.7^{-1.3}_{+0.9}$ & $20^{+3}_{-5}$\\

\multicolumn{3}{l}{(1s$^{2}$2p$^{5}$-2p$^{4}$($^{1}$D)3d)} & ~ & ~ & ~ & ~ & ~ & ~ & ~ & ~ \\

\multicolumn{3}{l}{Fe XIX} &
14.53 & $14.510^{+0.006}_{-0.005}$ & $400^{+100}_{-120}$ & $5.5(1.6)$
& 1100(300) & $-15.2^{-3.8}_{+3.0}$ & $-10.6^{02.7}_{+2.1}$ & $530^{+140}_{-110}$\\

\multicolumn{3}{l}{(2s$^{2}$2p$^{4}$-2p$^{3}$($^{2}$P)3s)} & ~ & ~ & ~ & ~ & ~ & ~ & ~ & ~  \\

\multicolumn{3}{l}{Fe XVIII} &
14.610 & $14.613^{+0.004}_{-0.005}$ & $-60^{-80}_{+100}$ &
$3.2^{+2.0}_{-0.1}$ & $650^{+350}_{-20}$ & $-13.7^{-5.3}_{+0.7}$ &
$-9.4^{-3.6}_{+0.4}$ & $480^{+180}_{-20}$\\

\multicolumn{3}{l}{(2s$^{2}$2p$^{5}$-2p$^{4}$($^{3}$P)3d)} & ~ & ~ & ~ & ~ & ~ & ~ & ~ & ~ \\

\multicolumn{3}{l}{Fe XIX} &
14.97 & $14.97^{+0.02}_{-0.01}$ & $100^{+160}_{-240}$ & 7.1 & 1400 &
$-22(3)$ & $-14.5(1.8)$ & 290(4) \\

\multicolumn{3}{l}{(2s$^{2}$2p$^{4}$-2p$^{3}$($^{4}$S)3s)} & ~ & ~ & ~ & ~ & ~ & ~ & ~ & ~ \\

\multicolumn{3}{l}{Fe XVIII} & 16.012 & $16.016(8)$ &
$80^{+140}_{-160}$ & $3.1^{+1.3}_{-1.0}$ & $580^{+240}_{-190}$ &
$-1.3^{-0.5}_{+0.4}$ & $-5.6^{-2.0}_{+1.5}$ & $5.8^{+2.2}_{-1.8}$\\

\multicolumn{3}{l}{(2s$^{2}$2p$^{5}$-2p$^{4}$($^{3}$P)3s)} & ~ & ~ & ~ & ~ & ~ & ~ & ~ & ~ \\

\tableline
\end{tabular}
\vspace*{\baselineskip}~\\ \end{center} \tablecomments{Parameters
obtained by fitting simple Gaussians to prominent line features in the
6--15~$\AA$ region.  1$\sigma$ errors are quoted; where errors are
not given, the parameter is fixed.  Errors in parentheses reflect
symmetric error values; single-digits in parentheses are the error in
the last digit of the measured value.  Column densities (N$_{\rm Z}$)
were calculated using oscillator strengths reported by Verner et
al. (1996) and Mewe et al (1985).  Theoretical wavelength values are
also taken from these references.  Negative equivalent width and flux
values indicate lines detected in absorption.\\ $\dag$ denotes a line
for which the wavelength was fixed; flux and equivalent width values
are effectively upper-limits.\\ $\dag\dag$ denotes a line for which
wavelength was fixed and the intensity was tied to the $r$ line
intensity; the flux and equivalent width values are effectively
upper-limits.\\ }
\vspace{-1.0\baselineskip}
\end{scriptsize}
\end{table}

\begin{figure}
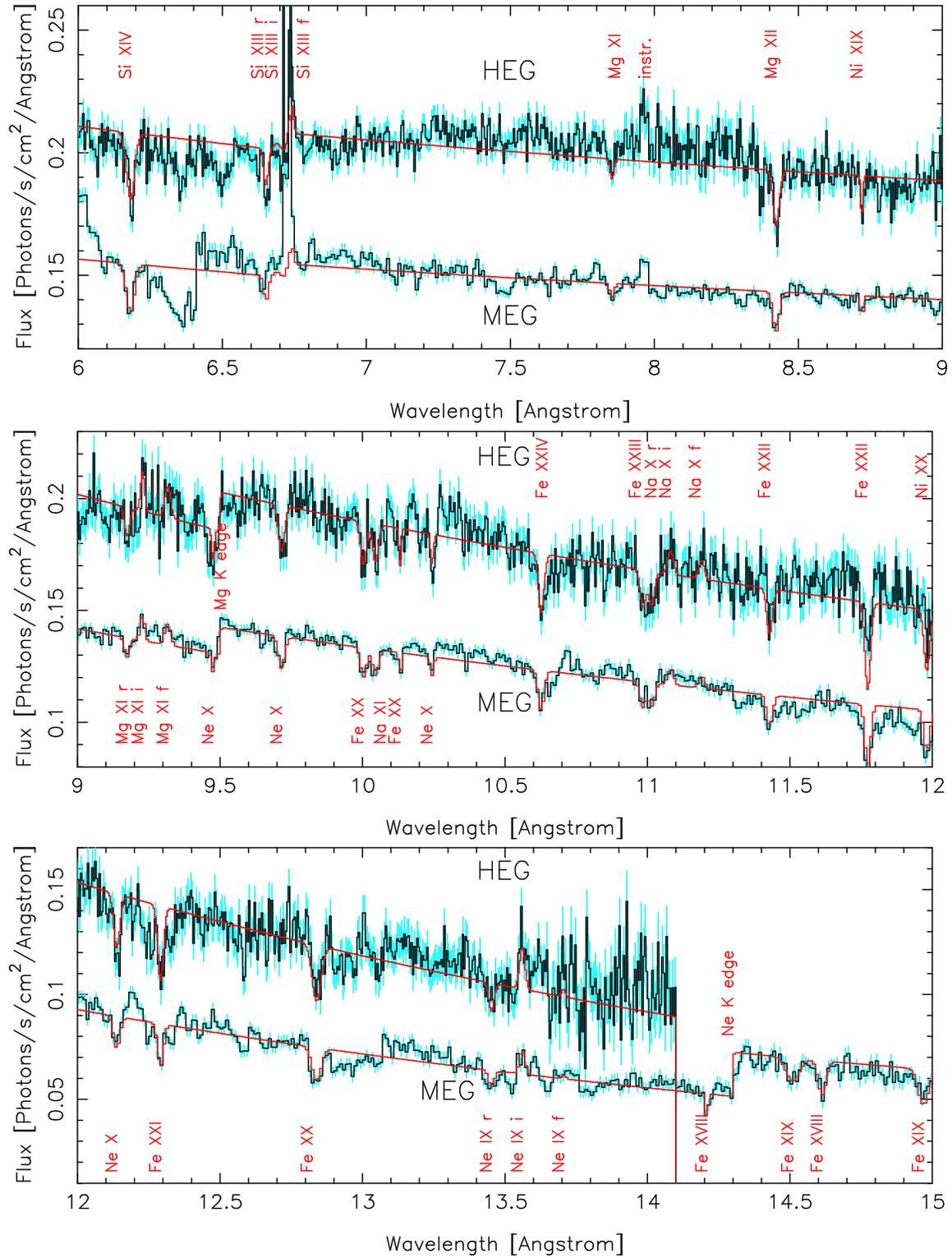

\figurenum{1}
\label{fig:six-fifteen}
\centerline{~\psfig{file=f1.ps,width=6.0in}~}
\centerline{~\psfig{file=f2.ps,width=6.0in}~}
\centerline{~\psfig{file=f3.ps,width=6.0in}~}
\caption{\footnotesize The highly-ionized absorption-dominated
spectrum.  Most lines are consistent with zero velocity shift, and a
FWHM of $\sim$700~km/s.  We interpret these features as absorption in
the companion wind.  Here, we fit only the strongest features seen in
both HEG and MEG orders.  There is clearly a myriad of weaker lines in
this spectrum.  The notch between 6-6.5\AA~ in the MEG is due to a
chip gap; the single-bin (in the HEG) ``emission'' feature at 6.71\AA~
and the broader feature at 7.96\AA~ are instrumental artifacts.  The
HEG is offset by $+0.04$ for visual clarity, and does not cover
wavelengths above $\sim14$\AA.  Please refer to Table 1 for the line
parameters and identifications.}
\end{figure}

\clearpage

\begin{figure}
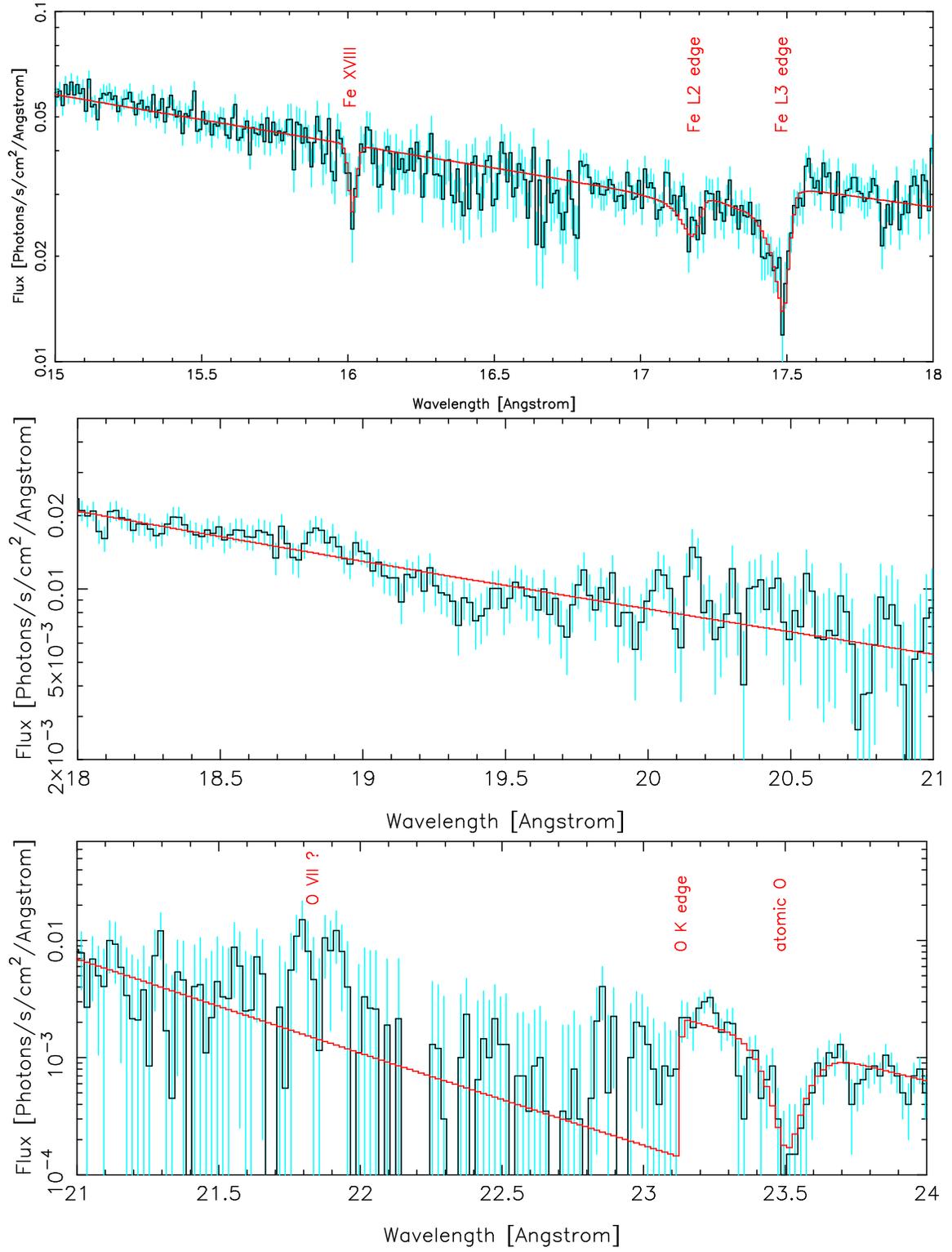

\figurenum{2}
\label{fig:six-fifteen}
\centerline{~\psfig{file=f4.ps,width=6.0in}~}
\centerline{~\psfig{file=f5.ps,width=6.0in}~}
\centerline{~\psfig{file=f6.ps,width=6.0in}~}
\caption{\footnotesize The MEG spectrum in the 15--24\AA~range.  As
the sensitivity in this range is lower, fewer features are clearly
detected.  The spectrum in the top panel is shown at full resolution
(0.01\AA), and the bottom two are rebinned (0.02\AA).}
\end{figure}

\clearpage

\begin{figure}
\figurenum{3}
\label{fig:wind_pic}
\centerline{~\psfig{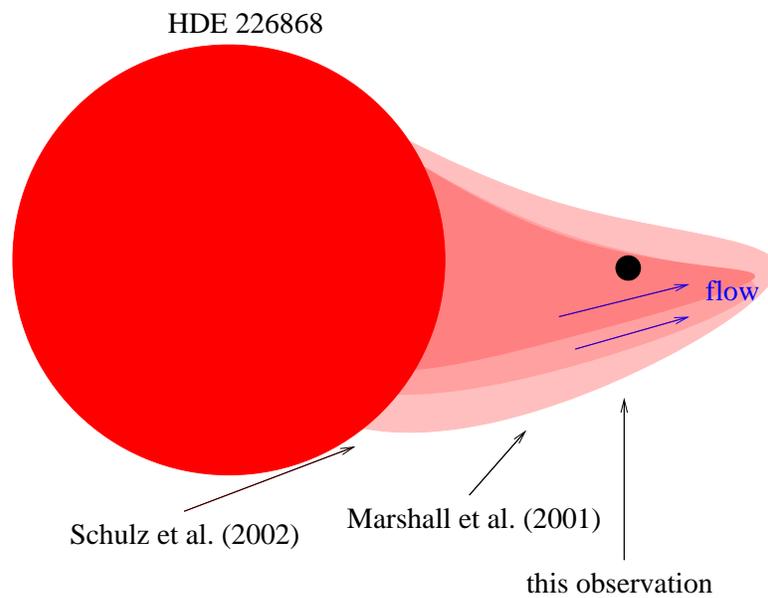}~}
\caption{\small A depiction of how the companion (HDE~226868,
O9.7~Iab) wind might focus near to the black hole in Cygnus X-1.  This
picture is based primarily on the Ne, Na, Mg, and Si Ly-$\alpha$ and
helium-like resonance absorption line velocities measured in this
observation and previously by Marshall et al. (2001), and possible
P-Cygni-type line profiles reported by Schulz et al. (2002).  Note
that this picture is a projection in the $\phi$ plane and that the
system is likely observed at $\theta \simeq 35^{\circ}$ (Gies \&
Bolton 1986).  At larger radii, a radial wind flow component is likely
important, but is not shown to emphasize a possible internal
geometry.}
\end{figure}

\end{document}